\documentstyle[preprint,aps,epsfig]{revtex}
\begin{document}
\title{Probing the matter term at long baseline experiments}
\author
{Mohan Narayan and S. Uma Sankar}  
\address
{Department of Physics, I.I.T. , Powai, Mumbai 400076, India}
\date{\today}
\maketitle

\begin{abstract}
We consider $\nu_\mu \rightarrow \nu_e$ oscillations in long
baseline experiments within a three flavor
oscillation framework. A non-zero measurement of this oscillation
probability implies that the $(13)$ mixing angle $\phi$ is non-zero. 
We consider the effect of
neutrino propagation through the matter of earth's crust and show that, 
given the constraints from solar neutrino and CHOOZ data, matter effects 
enhance the mixing for neutrinos rather than for anti-neutrinos. 
We need data from two different
experiments with different baseline lengths (such as K2K and MINOS) to 
distinguish matter effects unambiguously. 
\end{abstract} 
\vspace{0.5cm}

Recent results from the Super-Kamiokande have sparked tremendous interest
in neutrino physics \cite{skan,sksn}. The deficits seen in the solar and
atmospheric neutrino fluxes can be very naturally explained in terms of
neutrino oscillations. Since the energy and the distance scales of the
solar neutrino problem are widely different from those of the atmospheric
neutrino problem, the neutrino oscillation solution to these problems
requires widely different values of mass-squared differences 
\cite{nmru,flm}. 
A minimum of three mass eigenstates, and hence three neutrino flavors,
are
needed for a simultaneous solution of solar and atmospheric neutrino
problems. Since LEP has shown that three light, active neutrino species
exist \cite{lep}, it is natural to consider all neutrino data in the
framework of oscillations between all the three active neutrino flavors.

The flavor eigenstates $\nu_\alpha  (\alpha = e, \mu, \tau)$ are related
to the mass eigenstates $\nu_i (i = 1,2,3)$ by a unitary matrix $U$
\begin{equation}
|\nu_\alpha \rangle = U | \nu_i \rangle. \label{eq:flmarel}
\end{equation}
As in the quark sector, $U$ can be parametrized in terms of three mixing
angles and one phase. A widely used parametrization, convenient for
analyzing neutrino oscillations with matter effects, is \cite{kupa} 
\begin{equation} 
U = U^{23} (\psi) \times U^{phase} \times U^{13} (\phi)
\times U^{12} (\omega),
\label{eq:defU}
\end{equation}
where $U^{ij} (\theta_{ij})$ is the two flavor mixing matrix between 
the $i$th and $j$th mass eigenstates with the mixing angle $\theta_{ij}$.
We assume that the vacuum mass eigenvalues have the pattern $\mu_3^2 \gg
\mu_2^2 >
\mu_1^2$. Hence $\delta_{31} \simeq \delta_{32} \gg \delta_{21}$, where
$\delta_{ij} = \mu_i^2 - \mu_j^2$. The larger $\delta$ sets the scale for
the
atmospheric neutrino oscillations and the smaller one for solar
neutrino oscillations. The angles $\omega, \phi$ and $\psi$ vary in the
range $\left[0, \pi/2\right]$. For this range of the mixing angles, there
is no loss of generality due to the assumption that $\delta_{31},
\delta_{21} \geq 0$ \cite{flms}.

It has been shown \cite{kupa,jim94} that in the above approximation of one
dominant mass, the oscillation probability for solar neutrinos is a
function of only three parameters ($\delta_{21}, \omega$ and $\phi$) and
that of the atmospheric neutrinos and long baseline neutrinos is also
function of only three
parameters ($\delta_{31}, \phi$ and $\psi$). In each case, the three
flavor nature of the problem is illustrated by the fact that the
oscillation probability is a function of two mixing angles. 
The phase is unobservable in the one dominant mass approximation because,
in both solar and atmospheric neutrino oscillations, one of the three 
mixing angles can be set to zero \cite{flms}.  
The value of $\delta_{31}$ preferred by the Super-K
atmospheric neutrino data is about $2 \times 10^{-3}$ eV$^2$ 
\cite{skan}. For this
value of $\delta_{31}$, CHOOZ data sets a very strong constraint
\cite{choozex,choozppr} 
\begin{equation}
\sin^2 2 \phi \leq 0.2. \label{eq:chzcnst}
\end{equation}
If $\phi = 0$, then the solar and atmospheric neutrino oscillations get
decoupled and become two flavor oscillations with relavant parameters
being $(\delta_{21}, \omega)$ and $(\delta_{31}, \psi)$ respectively.

It is interesting to look for the consequences of non-zero $\phi$. A 
non-zero $\phi$ leads to $\nu_\mu \rightarrow \nu_e$ oscillations in 
atmospheric neutrinos and long baseline experiments. 
Due to theoretical
uncertainty in the calculation of atmospheric neutrino fluxes, it will be
very difficult to discern the effect of a small $\phi$ from the
atmospheric neutrino data. Recently a proposal was made to look 
for matter enhanced $\nu_\mu \rightarrow \nu_e$ oscillations in 
atmospheric neutrino data, which can be significant even if $\phi$ 
is small \cite{jim98}. Here we consider the effect of matter on
$\nu_\mu \rightarrow \nu_e$ oscillations in a three flavor framework
at long baseline experiments. 

The relation between the flavor states and mass eigenstates can be 
written in the simple form, 
\begin{equation}
\left[ \begin{array}{c} \nu_e \\ c_\psi \nu_\mu - s_\psi \nu_\tau \\
s_\psi \nu_\mu + c_\psi \nu_\tau \end{array} \right] =
\left( \begin{array}{ccc} c_\phi & 0 & s_\phi \\ 0 & 1 & 0 \\
-s_\phi & 0 & c_\phi \end{array} \right) \left[ \begin{array}{c}
\nu_1 \\ \nu_2 \\ \nu_3 \end{array} \right], \label{eq:matrix}
\end{equation} 
where $c$ stands for cosine and $s$ stands for sine.  
This equation is obtained by first setting $U^{phase} = I$ and 
$\omega = 0$ in equation~(\ref{eq:defU}). Then substitute the
resulting form $U$ in equation~(\ref{eq:flmarel}) and multiply 
it from left by $U^{23} (-\psi)$.  
From~(\ref{eq:matrix}) we see that $\nu_2$ has no
$\nu_e$ component and hence is decoupled from any oscillation 
involving $\nu_e$. Thus the calculation of oscillation probability
is essentially a two flavor problem. The three flavor nature is
present in the fact that the oscillations occur between $(\nu_e
\rightarrow (s_\psi \nu_\mu + c_\psi \nu_\tau))$. 
The vacuum oscillation probability is calculated to be 
\begin{equation}
P_{\mu e} = \sin^2 \psi \sin^2 2 \phi \sin^2 \left( \frac{1.27  \
\delta_{31}
L}{E} \right), \label{eq:vacP}
\end{equation}
where $\delta_{31}$ is in eV$^2$, the baseline length $L$ is in km and 
the neutrino energy $E$ is in GeV. For low energies the phase $1.27 
\delta_{31} L/E$ oscillates rapidly and $P_{\mu e}$ becomes insensitive
to it. However, for the range of energies for which the phase is close
to $\pi/2$, $P_{\mu e}$ varies slowly as shown in figures (1)-(3).
If the spectrum of the neutrinos in long baseline experiments peaks in
this range, then $\delta_{31}$ and the mixing angles can be 
determined quite accurately.  

Given the CHOOZ constraint on
$\phi$, Super-Kamiokande atmospheric neutrino data 
fix $\sin^2 2 \psi \simeq 1$ or $\sin^2 \psi \simeq 0.5$.
Substituting this value in equation~(\ref{eq:vacP}),  
we find that $P_{\mu e} \leq 0.1$. The neutrino beams at K2K
and MINOS have a $1 \%$ $\nu_e$ contamination. This background
and the systematic uncertainties set a limit to the smallest 
value of $P_{\mu e}$ (and $\phi$) that can be measured.  
With the current estimates of the systematic uncertainties,
the sensitivity of K2K is similar to that of CHOOZ ($\sin^2 2 \phi
\sim 0.2$) but MINOS is capable of measuring values of $\phi$ as 
small as $3^o (\sin^2 2 \phi \geq 0.01)$ \cite{oyama}. Once K2K
starts running, its systematics will be better understood and 
its sensitivity is likely to improve.

In both K2K and MINOS experiments, the neutrino beam travels through
earth's crust, where it encounters matter of roughly constant density 
3 gm/cc. This traversal leads to the addition of the Wolfenstein term to
the $e-e$ element of the (mass)$^2$ matrix, when it is written in the
flavor basis \cite{wolf}. The Wolfenstein term is given by
\begin{equation}
A = 0.76 \times \rho ({\rm gm/cc}) \times E ({\rm GeV}) \times 10^{-4} \
{\rm eV}^2.
\end{equation}
For $\rho$ of a few gm/cc and $E$ of a few GeV, $A$ is comparable to
the value of $\delta_{31}$ set by atmospheric neutrino data. 
This can lead to
interesting and observable matter effects in long baseline experiments.
The interactions of $\nu_\mu$ and $\nu_\tau$ with ordinary matter 
are identical. Hence equation~(\ref{eq:matrix}) can be used to 
include matter effects in a simple manner because the  
problem, once again, is essentially a two flavor one. 
It is easy to see that $\psi$ is unaffected by
matter and the matter dependent mixing angle $\phi_m$ and the mass
eigenvalues $m_1^2$ and $m_3^2$ are given by \cite{jim94} 
\begin{eqnarray}
\tan 2 \phi_m & = & \frac{\delta_{31} \sin 2 \phi}{
\delta_{31} \cos 2 \phi - A}, \label{eq:tan2phim} \\
m_1^2 & = &  \frac{1}{2} \left[ \left(\delta_{31} + A \right)
           - \sqrt{\left(\delta_{31} \cos 2 \phi - A \right)^2 +
                   \left(\delta_{31} \sin 2 \phi \right)^2} \right] , 
\label{eq:m1sq}  \\
m_3^2 & = &  \frac{1}{2} \left[ \left(\delta_{31} + A \right)
           + \sqrt{\left(\delta_{31} \cos 2 \phi - A \right)^2 +
                   \left(\delta_{31} \sin 2 \phi \right)^2} \right] . 
\label{eq:m3sq}  
\end{eqnarray}
The above equations hold for the propagation of neutrinos. For vacuum
propagation, the mass eigenvalues and mixing angles for neutrinos and
anti-neutrinos are the same. However, to include matter effects for
anti-neutrinos, one should replace $A$ by $-A$ in
equations~(\ref{eq:tan2phim}), (\ref{eq:m1sq}) and~(\ref{eq:m3sq}).
Note that, since $A$ is always positive, matter effects enhance the 
neutrino mixing angle and suppress the anti-neutrino mixing angle if
$(\delta_{31} \cos 2 \phi)$ is positive and vice-verse if $(\delta_{31}  
\cos 2 \phi)$ is negative. Since we have taken $\delta_{31}$
to be positive, matter effects enhance neutrino mixing if $\cos 2 \phi$ is
positive or $\phi < \pi/4$. The CHOOZ constraint from
equation~(\ref{eq:chzcnst}) on $\sin^2 2 \phi$ sets
the limit $\phi \leq 13^o$ or $\phi \geq 77^o$. For the latter
possibility, $\cos 2 \phi$ is negative. However, this large a value of
$\phi$ is forbidden by the three flavor analysis of solar neutrino data,
which yields the independent
constraint $\phi \leq 50^o$ \cite{nmru,flm}. Hence the enhancement of the
mixing angle will be for neutrinos rather than for anti-neutrinos. This is
good news because the beams in long baseline experiments  
consist of neutrinos overwhelmingly.

The matter dependent oscillation probability can be calculated in
a straight forward manner from equation~(\ref{eq:matrix}) in terms of 
$\psi, \phi_m$ and $\delta^m_{31} = m^2_3 - m^2_1$.
\begin{equation}
P^m_{\mu e} = \frac{1}{2} \sin^2 2 \phi_m \sin^2 \left( \frac{1.27
\delta^m_{31} L}{E} \right), \label{eq:matP}
\end{equation}
where we have substituted the Super-Kamiokande best fit value $\sin^2 \psi
= 1/2$. As discussed above, the matter effects enhance the mixing angle
for neutrinos and the matter modified oscillation probability $P^m_{\mu
e}$ will be greater than the vacuum oscillation probability $P_{\mu e}$,  
for a range of energies. For some neutrino energy, the
Mikheyev-Smirnov resonance condition \cite{miksmi} 
\begin{equation}
\delta_{31} \cos 2 \phi = A \label{eq:rescon}
\end{equation}
can be satisfied and near this energy $\phi_m \sim \pi/4$. Unfortunately,
this does not lead to any dramatic increase in $P^m_{\mu e}$ because, near
the resonance, when $\sin^2 2 \phi_m$ is maximized, the matter dependent
mass-squared difference $\delta^m_{31}$ is minimized \cite{ajmm}. 
In fact, near the 
resonance, $P^m_{\mu e} \geq P_{\mu e}$. By differentiating
equation~(\ref{eq:matP}) with respect to $E$, we can calculate the energy
at which $P^m_{\mu e}$ is maximized. The highest energy at which this
occurs is just below the highest energy at which $P_{\mu e}$ is maximised,
that is the energy for which the phase $1.27 \delta_{31} L/E = \pi/2$.
At this energy, the $\phi_m$ is significantly higher than $\phi$ and
hence $P^m_{\mu e}$ will be measurably higher than $P_{\mu e}$. 

Since the baseline of K2K is $3$ times smaller than that of MINOS, the 
energy where the its phase becomes $\pi/2$ is smaller by a factor of
$3$ compared to similar energy for MINOS. Because the highest energy
maximum (at which the phase is $\pi/2$) occurs at different energies, 
the increase in $\nu_\mu 
\rightarrow \nu_e$ oscillation probability, due to the energy dependent
enhancement of the mixing angle, is different for the two experiments. 
We find that, at their respective highest energy maxima,  
$P^m_{\mu e} \simeq 1.1 P_{\mu e}$ for K2K
and $P^m_{\mu e} = 1.25 P_{\mu e}$ for MINOS. This is illustrated
in figure~(1) for $\phi \simeq 12.5^o$ (which is just below the CHOOZ 
limit) and $\delta_{31} = 2 \times 10^{-3}$ eV$^2$ (which is the
best fit value for Super-Kamiokande atmospheric neutrino data).
This conclusion is
indenpendent of the value of $\delta_{31}$ and is illustrated in figures
(1)-(3), for different values of $\delta_{31}$ (and $\phi = 12.5^o$).
A similar conclusion was
obtained a recent paper, where it was demonstrated that the relation 
between $P^m_{\mu e}$ and $P_{\mu e}$ is almost independent of $\phi$
\cite{lipari}.

We wish to emphasize that data from at least two different experiments
with different baselines is needed to state unambiguously, whether matter 
effects are playing a role in neutrino oscillations.
Suppose we have data from only one experiment. We can obtain allowed
values of $\delta_{31}$ and $\phi$ by fitting either $P_{\mu e}$ or 
$P^m_{\mu e}$ to this data. The two analyses will give 
different values of mixing angle. Since
we don't apriori know the value of vacuum mixing angle, we can't say 
which result is correct. However, as mentioned above, matter effects
lead to different enhancement of oscillation probability for K2K and
MINOS as shown in figure~(1). This difference can be exploited to 
distinguish matter effects. The data from each experiment, K2K and
MINOS, should be analyzed twice, once using $P_{\mu e}$ as the input 
and the second time with $P^m_{\mu e}$ as the input. In the first
analysis, the allowed value of the mixing angle from MINOS will be
significantly higher that from K2K, if the matter effects are 
important. Matter effects can be taken to be established, if the
second analysis gives the same values of $\phi$ for both K2K and MINOS.

In conclusion, we find that matter effects enhance $\nu_\mu
\rightarrow \nu_e$ oscillations at long baseline experiments 
K2K and MINOS. This enhancement can be large enough to be 
observable. However, one must combine the data from both 
experiments before making a definite statement about the 
effect of the matter term.

Acknowledgement: We thank Sameer Murthy for his help in the preparation 
of the figures.

\newpage

\begin{figure}
{\centering \includegraphics{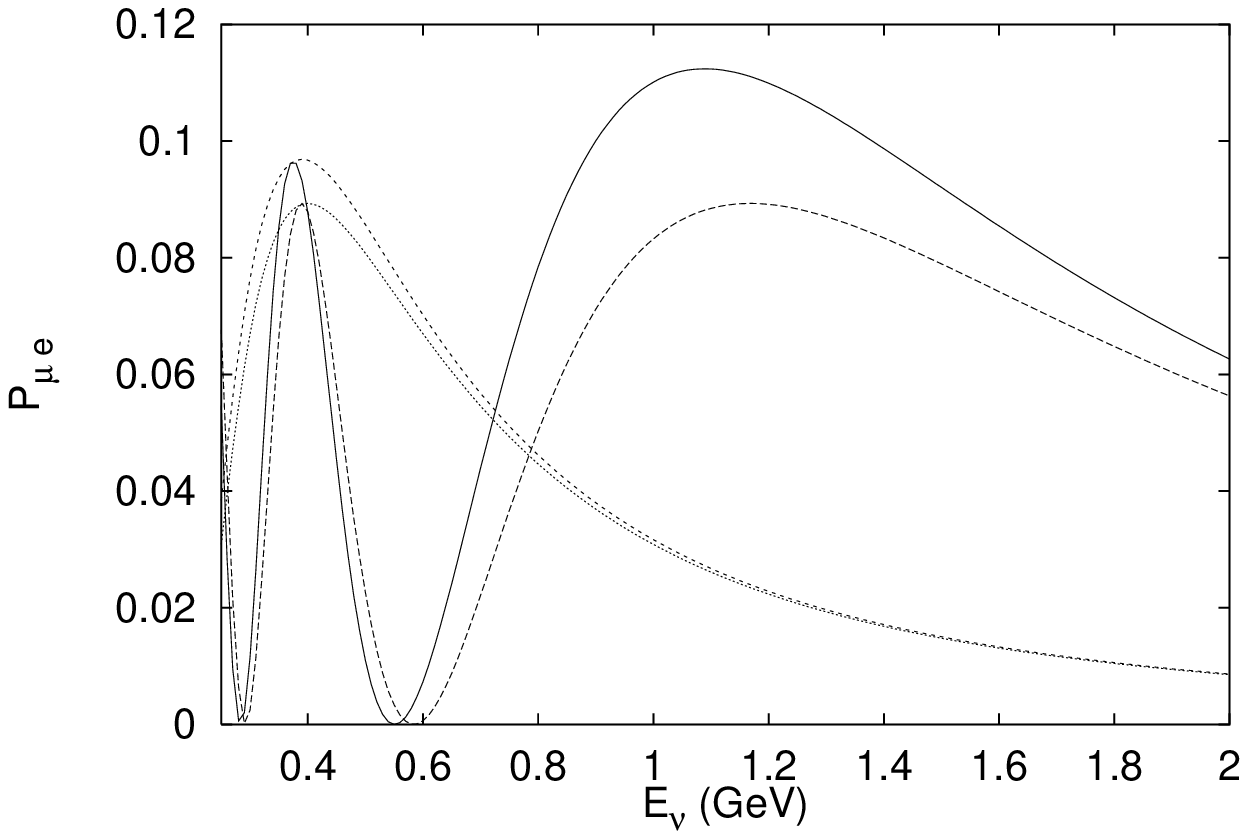} \par}
\caption{
Plots of $P^m_{\mu e}$ for MINOS (solid line),
$P_{\mu e}$ for MINOS (long-dashed line), 
$P^m_{\mu e}$ for K2K (short-dashed line),
$P_{\mu e}$ for K2K (dotted line) vs $E$ for  
$\delta_{31} = 2 \times 10^{-3}$ eV$^2$ and $\phi = 12.5^o$.} 
\end{figure}

\begin{figure}
{\centering \includegraphics{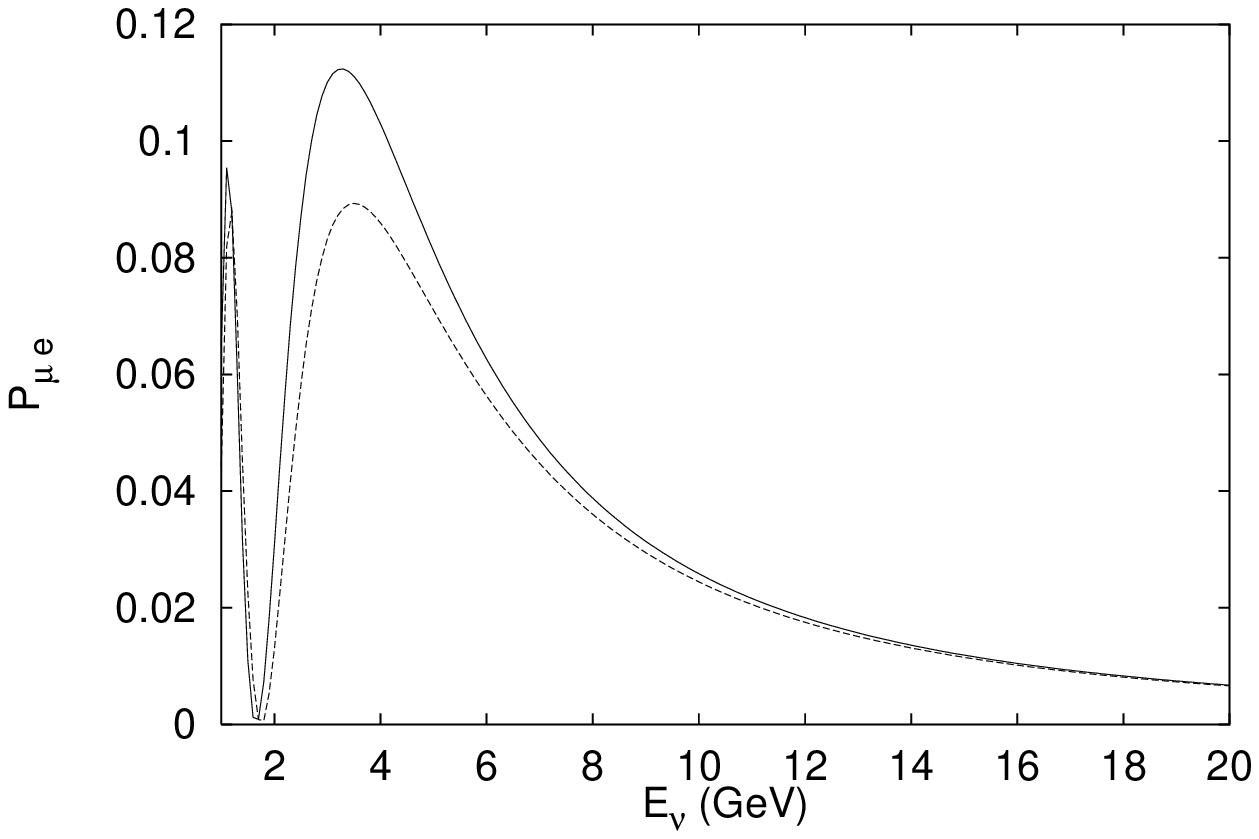} \par}
\caption{ 
Plots of $P^m_{\mu e}$ (solid line),
$P_{\mu e}$ (long-dashed line) vs $E$ for MINOS  
$\delta_{31} = 6 \times 10^{-3}$ eV$^2$. and $\phi = 12.5^o.$}
\end{figure}

\begin{figure}
{\centering \includegraphics{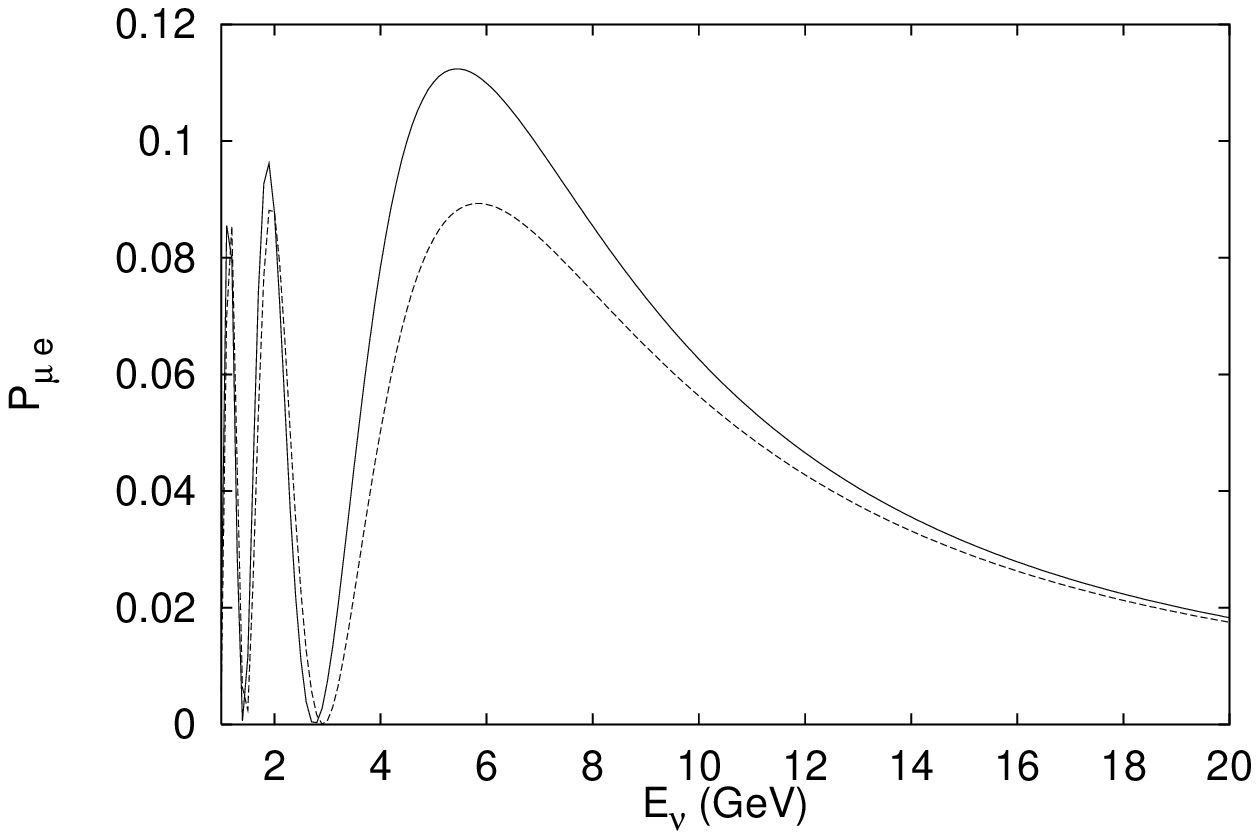} \par}
\caption{ 
Plots of $P^m_{\mu e}$ (solid line),
$P_{\mu e}$ (long-dashed line) vs $E$ for for MINOS  
$\delta_{31} = 10^{-2}$ eV$^2$. and $\phi = 12.5^o.$}
\end{figure}


\begin{thebibliography}{abcdefghhijk} 
\bibitem{skan}
Super-Kamiokande Collaboration: Y. Fukuda {\it et al}, Phys. Rev. Lett.
{\bf 81}, 1562 (1998). 
\bibitem{sksn}
Super-Kamiokande Collaboration: Y. Fukuda {\it et al}, Phys. Rev. Lett.
{\bf 82}, 1810 (1999).
\bibitem{nmru}
M. Narayan, M. V. N. Murthy, G. Rajasekaran and S. Uma Sankar, Phys. Rev.
{\bf D 53}, 2809 (1996).
\bibitem{flm}
G. Fogli, E. Lisi and D. Montanino, Phys. Rev. {\bf D 54}, 2048 (1996).
\bibitem{lep}
Particle Data Group, European Physics Journal, {\bf 3}, 23 (1998). 
\bibitem{kupa}
T. K. Kuo and J. Pantaleone, Rev. Mod. Phys. {\bf 61}, 937 (1989).
\bibitem{flms}
G. Fogli, E. Lisi, D. Montanino and G. Scoscia, Phys. Rev. {\bf D 55}, 
4385 (1997), especially Apprendix C.
\bibitem{jim94}
J. Pantaleone, Phys. Rev. {\bf D49}, R2152 (1994).
\bibitem{choozex}
CHOOZ Collaboration: M. Appolonio {\it et al}, Phys. Lett. {B 420}, 
397 (1998).
\bibitem{choozppr}
M. Narayan, G. Rajasekaran and S. Uma Sankar, Phys. Rev. {\bf D58},
R031301 (1998).
\bibitem{jim98}
J. Pantaleone, Phys. Rev. Lett. {\bf 81}, 5060 (1998).
\bibitem{oyama}
Y. Oyama, Talk presented at YITP workshop on flavor physics, Kyoto,
January 1998; Preprint hep-ex/9803014. 
\bibitem{wolf}
L. Wolfenstein, Phys. Rev. {\bf D 17}, 2369 (1978); {\it ibid} 
{\bf D 20}, 2634 (1979).
\bibitem{miksmi}
S. P. Mikheyev and A. Yu. Smirnov, Yad. Fiz. {\bf 42}, 1441 (1985)
[Sov. J. Nucl. Phys. {\bf 42}, 913 (1985)]; Nuovo Cimento {\bf C9},
17 (1986). 
\bibitem{ajmm}
A. S. Joshipura and M. V. N. Murthy, Phys. Rev. {\bf D 37}, 
1374 (1988).
\bibitem{lipari}
P. Lipari, hep-ph/9903541
\end{thebibliography}
\end{document}